\documentclass{scrartcl}

\usepackage{amsmath, amssymb}
\usepackage{hyperref}
\RequirePackage[final]{graphicx}

\DeclareOldFontCommand{\bf}{\normalfont\bfseries}{\mathbf}

\begin{document}

\author{Mario~Schwarz\footnote{Corresponding author\newline Email address: mario.schwarz@tum.de}, Sabrina~M.~Franke, Lothar~Oberauer, \\ Miriam~D.~Plein, Hans~Th.~J.~Steiger and Marc~Tippmann \and \\
{\large \textit{Physik-Department, Technische Universit\"{a}t M\"{u}nchen}} \\ {\large \textit{James-Franck-Str. 1, 85748 Garching, Germany}}}

\title{Measurements of the Lifetime of Orthopositronium in the LAB-Based Liquid Scintillator of JUNO}

\maketitle















\begin{abstract}
\noindent
\textsc{Abstract:} Electron antineutrinos are detected in organic liquid scintillator based neutrino experiments by means of the inverse beta decay, producing both a positron and a neutron. The positron may form a bound state together with an electron, called positronium (Ps). The longer-lived spin state of Ps, orthopositronium (o-Ps) has a lifetime of about $3\,\mathrm{ns}$ in organic liquid scintillators (LS). Its formation changes the time distribution of photon emission, which affects positron reconstruction algorithms and allows the application of pulse shape discrimination (PSD) to distinguish electron from positron events. \\
In this work, we measured the lifetime $\tau_2$ of o-Ps in the linear alkylbenzene (LAB) based LS of the JUNO (Jiangmen Underground Neutrino Observatory) experiment including wavelength shifters, obtaining $\tau_2 = 2.97\,\mathrm{ns} \pm 0.04\,\mathrm{ns}$. Due to systematics, which are not yet completely understood, we are not able to give a final result for the o-Ps formation probability $I_2$. We use a novel type of setup, which allows a better background suppression as compared to commonly used PALS (positron annihilation lifetime spectroscopy) measurements. \\

\end{abstract}




\section{Introduction}



Low-energy electron antineutrino detection in liquid scintillator (LS) neutrino detectors has a long and successful history. Recently, the neutrino mixing angle $\Theta_{13}$ has been determined by the reactor neutrino experiments Double Chooz \cite{Abe:2015rcp}, Daya Bay \cite{dayaBay} and RENO \cite{Kim:2016yvm}. 
Furthermore, the planned JUNO (Jiangmen Underground Neutrino Observatory), a reactor electron antineutrino LS experiment, aims to determine the neutrino mass hierarchy (NMH) \cite{JUNO:cdr,JUNO:nphys}. \\
All the mentioned experiments detect electron antineutrinos by means of the inverse beta decay (IBD) process:
\begin{equation}
 \overline{\nu_e} + p \rightarrow e^+ + n
\end{equation} 
The positron deposits its kinetic energy quickly, leading to a prompt signal \cite{JUNO:nphys}. The neutron is thermalized and eventually captured, producing a gamma, which generates a second signal \cite{JUNO:nphys}. Neutron capture times are in the order of a few tenth up to hundreds of microseconds \cite{franco}. \\
After depositing their kinetic energies in the LS, positrons either annihilate with electrons from the surrounding me\-di\-um or form positronium (Ps), a bound state consisting of a positron and an electron \cite{PhysRev.82.455}. Two spin states of Ps exist: the singlet state, parapositronium (p-Ps), which constitutes 25\% of the formed Ps and has a lifetime of $125\,\mathrm{ps}$ in vacuum \cite{PhysRevLett.72.1632}; the triplet state, orthopositronium (o-Ps), has a lifetime of $142\,\mathrm{ns}$ in vacuum \cite{PhysRevLett.65.1344}. While p-Ps dominantly decays via the diametrical emission of two $511\,\mathrm{keV}$ gammas, for the dominant decay channel of o-Ps in vacuum 3 gammas are emitted. \\

o-Ps within materials is affected by different processes, for example \textit{pickoff} (annihilation of the positron of o-Ps with an electron of the surrounding medium \cite{tao}), \textit{conversion} (spin-flip of Ps \cite{PhysRev.110.1355}) or chemical reactions \cite{tao}. These processes result in a two-body decay of o-Ps, shortening the effective lifetime to values from about $100\,\mathrm{ps}$ in metals \cite{bell:graham} to about $100\,\mathrm{ns}$ in gases \cite{PhysRev.82.455}. Typical lifetimes of o-Ps are in the range of a few ns in organic liquids \cite{gcs}, which are used in most liquid scintillation neutrino detectors. Since these lifetimes are small compared to the lifetime of o-Ps in vacuum, the fraction of three-gamma decay of o-Ps in organic liquids can be neglected \cite{brandt:berko}. \\
Since the lifetime of p-Ps is very close to the lifetime of directly annihilating positrons (about $100\,\mathrm{ps}$ \cite{RevModPhys.60.701}), they are not distinguished in most experiments and hence together form the short-lived component appearing in positron lifetime annihilation spectroscopy (PALS) \cite{tao}. The effective lifetime and their combined formation probability are referred to as $\tau_1$ and $I_1$, respectively. The long-lived state originates from o-Ps, its lifetime and formation probability is called $\tau_2$ and $I_2$, respectively. \\
The formation of o-Ps induces a delay between the deposition of a positron's kinetic energy in the LS and the signal caused by the annihilation gammas. Thus, the photon emission time distribution (\textit{pulse shape}) is changed with respect to direct annihilation \cite{franco}, enabling the application of pulse shape discrimination (PSD) methods \cite{RANUCCI1998374}. \\
The implementation of PSD in order to dicriminate between electron and positron events in LS neutrino detectors to suppress background processes has been proposed \cite{franco} and successfully applied \cite{Bellini:2013lnn}. Electron antineutrino detectors utilizing the IBD process, can profit from a discrimination between IBD signals and $\beta-n$ decays of cosmogenic $^8$He and $^9$Li, which mimic the IBD signal \cite{franco}. In JUNO, these isotopes are expected to constitute the largest source of background \cite{JUNO:nphys}. Therefore, an application of the $e^+$/$e^-$ discrimination in JUNO has been proposed \cite{Cheng:2016ego}. \\
However, since the lifetimes of o-Ps in LS typically are around $3\,\mathrm{ns}$, which is comparable to the fast scintillation component of LS, an identification of o-Ps is difficult \cite{Consolati:2015bma}. It has been shown via simulation that in a spherical neutrino detector of $4\,\mathrm{m}$ radius filled with a pseudocumene (PC, C$_{9}$H$_{12}$) based LS ($\tau_2 = 3.12\,\mathrm{ns}$, $I_2 = 51.2\%$) the identification efficiency of o-Ps is at most 25\% \cite{franco}. Furthermore, the probability of correctly identifying o-Ps events depends on the positron's kinetic energy. This is due to the fact, that the composite signal is more likely to be wrongly identified as being caused by a single energy distribution, if one of the overlapping signals deposits  more energy than the other one. Therefore the maximum o-Ps identification probability of 25\% applies to a positron kinetic energy of about $1022\,\mathrm{keV}$ \cite{franco}. \\
Since the efficiency of identifying o-Ps events is low, the $e^+$/$e^-$ discrimination in $\overline{\nu_e}$ detectors is most useful when applied in cases where a clean low-statistics $e^+$ sample is needed, for example for background studies. \\
In $\nu_e$ detectors, the $e^+$/$e^-$ discrimination allows to reduce background originating from the $\beta^+$ decaying isotopes, as already used in Borexino for cosmogenic $^{11}$C \cite{Bellini:2013lnn}. \\
In any case an experimental determination of o-Ps properties is necessary in order to estimate the effect of the induced pulse shape distortions on $e^+$ reconstruction algorithms \cite{franco}. \\

Several previous experiments have measured the lifetimes and formation probabilities of o-Ps in various LS. Most of them \cite{franco, consolati, Cheng:2016ego} used the PALS setup, with the $\beta^+$ decay isotope $^{22}$Na as positron source. The source is encapsulated and immersed in the LS. Gammas are detected by two scintillation detectors typically placed orthogonal. The time difference between the emission of the positron (detection of a simultaneously emitted $1275\,\mathrm{keV}$ gamma) and the detection of one of the two $511\,\mathrm{keV}$ gammas is used to extract $\tau_2$ by the mentioned experiments. \\
PALS measurements record positron induced events regardless of the positron's interaction position. Thus, posi-trons interacting within the source or its encapsulation distort the lifetime spectrum. In order to reconstruct $I_2$ of the investigated LS, this effect has to be eliminated. For example, the fraction of positrons annihilating with an electron in the source encapsulation can be estimated by performing measurements with different thicknesses of the encapsulation \cite{franco}. Additionally, the deviations caused by positrons interacting within the source or its encapsulation can be accounted for by additional parameters in the final fit of the lifetime spectrum \cite{Cheng:2016ego}. \\
o-Ps properties have been measured for various LS solvents, e.g. PC, phenylxylylethane (PXE, C$_{16}$H$_{18}$) and linear alkylbenzene\footnote{The LAB used in JUNO is a mixture of different molecules, which can be expressed as C$_9$H$_{12} + ($CH$_2)_m$, $m = 7 - 10$ \cite{marrodan,JUNO:nphys}} (LAB, C$_{18}$H$_{30}$) \cite{franco}. Furthermore, the properties of o-Ps have been studied as a function of dopant concentration \cite{consolati} and for PC with an admixture of a wavelength shifter, showing significant deviations in both $\tau_2$ and $I_2$, with respect to pure PC \cite{franco}. \\
Since $\tau_2$ and $I_2$ only have been measured for pure LAB so far \cite{franco, Cheng:2016ego}, we measured both values for the complete LS planned to be used in JUNO. This LS is composed of LAB as solvent and $3\,\mathrm{g/l}$ 2,5-diphenyloxazole (PPO) as well as $15\,\mathrm{mg/l}$ 1,4-bis(omethlystyryl)-benzene (bisMSB) \cite{JUNO:cdr}. The lifetime measurement was performed using a novel setup concept, which is designed to minimize background caused by positrons interacting within the source or its encapsulation. \\


\section{Experimental Setup}

\begin{figure*}[t]
 \centering
 \includegraphics[width=12cm]{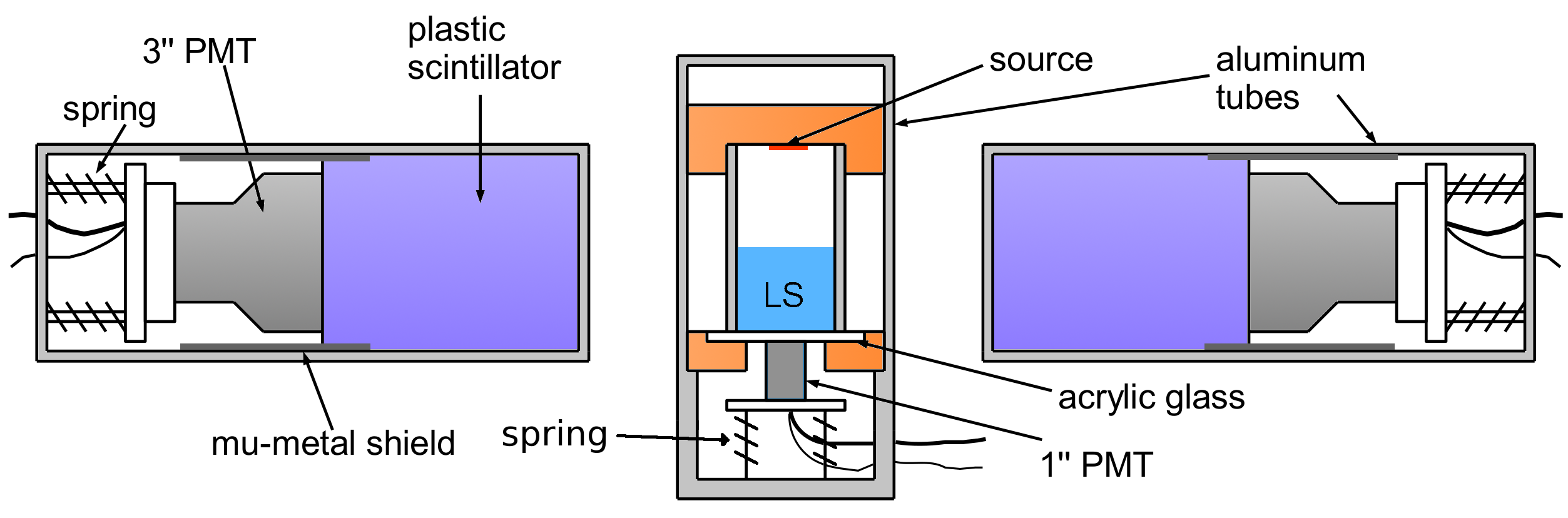} 
 \caption{ A sectional view of the setup used for the lifetime measurements. The two outer cylinders each contain a 3'' photomultiplier tube (PMT, rear) and a plastic scintillator (front). The central vessel is an upright cylinder, partially filled with liquid scintillator (LS) and containing the $^{22}$Na source, which is encapsulated in Kapton and fixed above the LS. A 1'' PMT is optically connected to the LS volume via an acrylic glass window.}
 \label{fig:setup}
\end{figure*}

A sectional view of the geometry of the setup is presented in figure \ref{fig:setup}. The setup consists of three aluminum tubes, acting as individual dark boxes. The central tube is in upright position and contains a cylindrical vessel, which is referred to as the central vessel in the following. The central vessel is partially filled with the LAB-based LS and also contains the $^{22}$Na positron source, encapsulated in Kapton \cite{kapton:sheet}. The source is mounted below the central vessel lid and there is no additional barrier between source and liquid scintillator. A 1'' photomultiplier tube (PMT) \cite{9111:sheet} is mounted at the bottom of the central tube and is pressed by springs to the acrylic glass window forming the bottom of the LS volume. \\
The outer two tubes are placed horizontally and contain a fast plastic scintillator (Bicron BC-400, $\tau = 2.4\,\mathrm{ns}$ \cite{bc400:sheet}) each in the front section. Springs press each a 3'' PMT \cite{9821:sheet} and a plastic scintillator together. The outer tubes and their interior are refered to as outer detectors in the following and they are focused towards the liquid level of the LS inside the central vessel. The whole setup is symmetrical with regard to the axis of the central vessel, hereinafter called the central axis. A dark box including a Faraday cage contains the whole geometry.\\
The measurement principle is based on positrons produced by $^{22}$Na hitting the LS volume and forming o-Ps with a certain probability. The lifetime of a single positron-induced event is obtained by measuring the time difference between a start and a stop signal. \\
The start signal consists of scintillation light produced by the positron depositing its kinetic energy in the LS, which is detected by the 1'' PMT. The duration of the energy deposition of the positron is of the order of $1\,\mathrm{ps}$ \cite{RevModPhys.60.701} and hence can be treated as being point-like in time. \\
Eventually, either by direct annihilation of the positron or by decay of the Ps, two $511\,\mathrm{keV}$ gammas are emitted diametrically. In case both gammas deposit energy in one outer detector each, a valid stop signal is obtained. \\
The geometry is chosen in such a way that background events are suppressed. The following aspects increase the background suppression compared to a PALS setup:
\begin{itemize}
 \item Positron induced signals are used for start signals in our measurement, as opposed to the $1275\,\mathrm{keV}$ gammas, used in PALS setups. This way, most events, where the positron does not hit the LS, are excluded in our experiment, as no valid start signal exists in this case. The only exception are $511\,\mathrm{keV}$ and $1275\,\mathrm{keV}$ gammas causing fake start signals. The fraction of events with a fake start signal is below 3\% according to the Monte Carlo based simulation described later. 
 \item Both $511\,\mathrm{keV}$ gammas are used as stop signals in our experiment, while only a single gamma is used in PALS experiments for that purpose. The detection of both gammas eliminates background consisting of single fake stop signals.
 \item The encapsulated positron source is not immersed in the LS, as opposed to PALS experiments. Therefore, annihilation gammas produced within the source and causing hits in both scintillation detectors are strong\-ly suppressed. This way, events where the positron annihilates in or near the source are sufficiently suppressed.
\end{itemize}

The major disadvantage of the preseted setup compared to a PALS setup is a lower acquisition rate, since less positrons reach the LS. \\
The geometry was implemented in a simulation based on the Monte Carlo (MC) framework Geant4 \cite{AGOSTINELLI2003250}. Single $^{22}$Na decays have been studied. The PENELOPE package \cite{penelope2008} was used for the simulation of electromagnetic interactions of electrons and positrons. Scintillation processes were also implemented in order to generate hit-time spectra for each of the three PMTs. \\
The Monte Carlo truth was used to classify events and different effects leading to triple coincidences\footnote{A single $^{22}$Na decay resulting in energy depositions in all scintillating volumes, i.e. with all PMTs being hit.} were identified and studied. \\
One type of triple coincidence corresponds to ``good'' events, from which the lifetime can be extracted in measurements. The positron has to deposit energy in the LS, annihilate inside of the LS and both $511\,\mathrm{keV}$ gammas have to interact with the outer detectors for the event to be classified as ``good''. Furthermore, the $1275\,\mathrm{keV}$ gamma, emitted simultaneously with the positron, may not hit any scintillating volume in ``good'' events and no single $511\,\mathrm{keV}$ gamma may hit both plastic scintillators. All these effects may cause fake triple coincidences. \\
Using the event classification, the geometry was optimized by varying two parameters of the simulated geometry: the distance between the outer detectors and the central axis (radial distance), as well as the filling height of the LS. A radial distance of $20\,\mathrm{cm}$ and a filling height of $55\,\mathrm{mm}$ was found to maximize the ratio of ``good'' events to all observed triple coincidences, while satisfying a minimum detection rate of ``good'' events of $100\,\mathrm{Hz}$. \\

The readout electronics of the experiment is based on a 4-channel-$2\,\mathrm{GS/s}$ FADC (Acqiris DC282 \cite{acqiris}), recording the signals produced by the three PMTs. A coincidence between both 3'' PMTs triggers the acqisition of an event. The data processing is done offline, as described in the next section. \\
In order to estimate the time resolution of the overall setup, a MC study based on measured PMT and FADC properties was performed. In this study the photon hit time spectra of all three channels, produced by the Geant4-based simulation were smeared with both transit time spectra (TTS) and single-photon pulse shapes of all three PMTs used. The TTS and the single-photon pulse shapes were obtained from single-photon measurements, as described in \cite{marc}. \\
By reconstructing the signal arrival times of the simulated waveforms, the lifetime\footnote{Only ``good'' events were used. Since the formation of Ps was not included in the Geant4-based simulation, the resulting lifetime spectrum only includes the detector time resolution.} of simulated events was obtained. The lifetime spectrum could be best described with the following detector resolution model of two gaussian distributions, where all parameters are free in the fit:
\begin{equation}
    \label{eq:detModel}
 f(t) = \rho \cdot \frac{1}{\sqrt{2 \pi \sigma_1^2}} \mathrm{e}^{-\frac{(t - \mu_1)^2}{2 \sigma_1^2}} + 
       (1 - \rho) \cdot \frac{1}{\sqrt{2 \pi \sigma_2^2}} \mathrm{e}^{-\frac{(t - \mu_2)^2}{2 \sigma_2^2}}
\end{equation} 
Here, $t$ denotes the lifetime, obtained by formula \ref{eq:lifetime}, and $\rho$ describes the ratio between both gaussian distributions. The relevant final fit parameters describing the detector resolution are given in the following table:
\begin{center}
\begin{tabular}{|l|l|l|l|l|l|}
\hline  & $\rho$ & $\sigma_1$ & $\sigma_2$  \\ 
\hline Value & 81.4\% & 0.377 ns & 1.258 ns \\ 
 Stat. error & 0.4\% & 0.002 ns & 0.012 ns \\ 
\hline 
\end{tabular}
\end{center}
Since only ``good'' events are used for this estimation, distortions caused by fake triple coincidences are expected to decrease the presented detector resolution.

\newpage

\section{Data Analysis and Results}

\begin{figure*}[t]
 \centering
 \includegraphics[width=12cm]{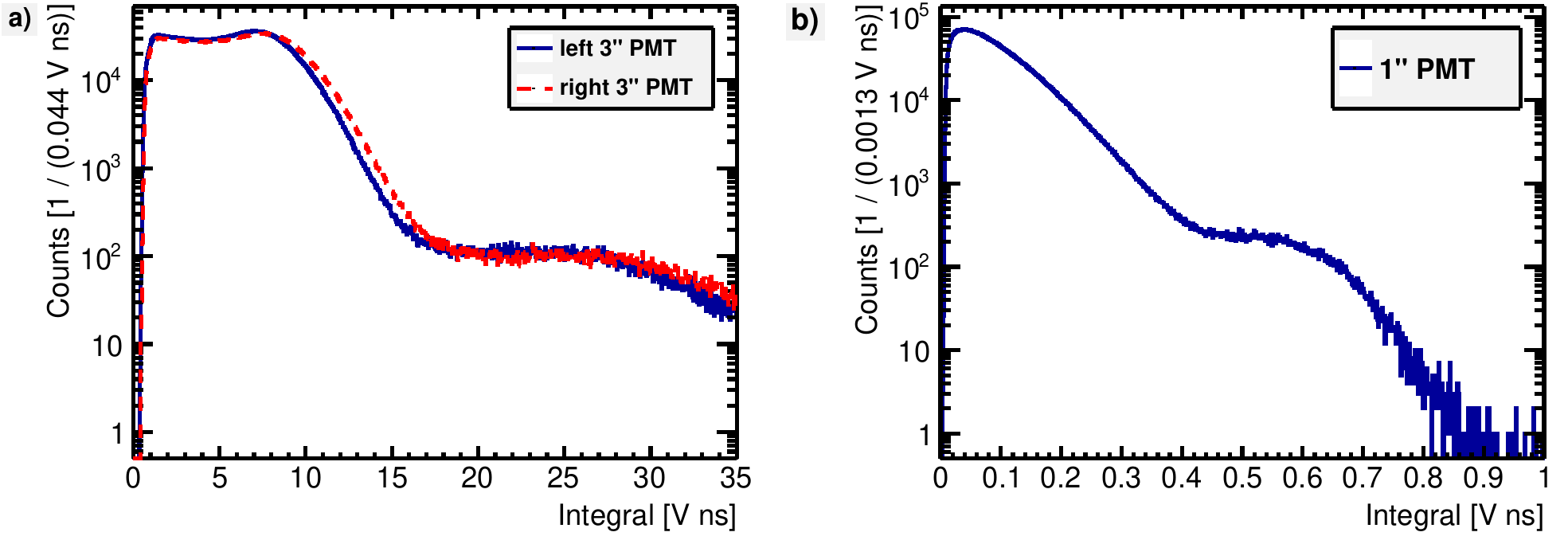} 
 \caption{ Measured energy spectra of triple coincidences for (a) both outer detectors and (b) the central 1'' PMT. }
 \label{fig:spec_meas}
\end{figure*}

\begin{figure*}[t]
 \centering
 \includegraphics[width=12cm]{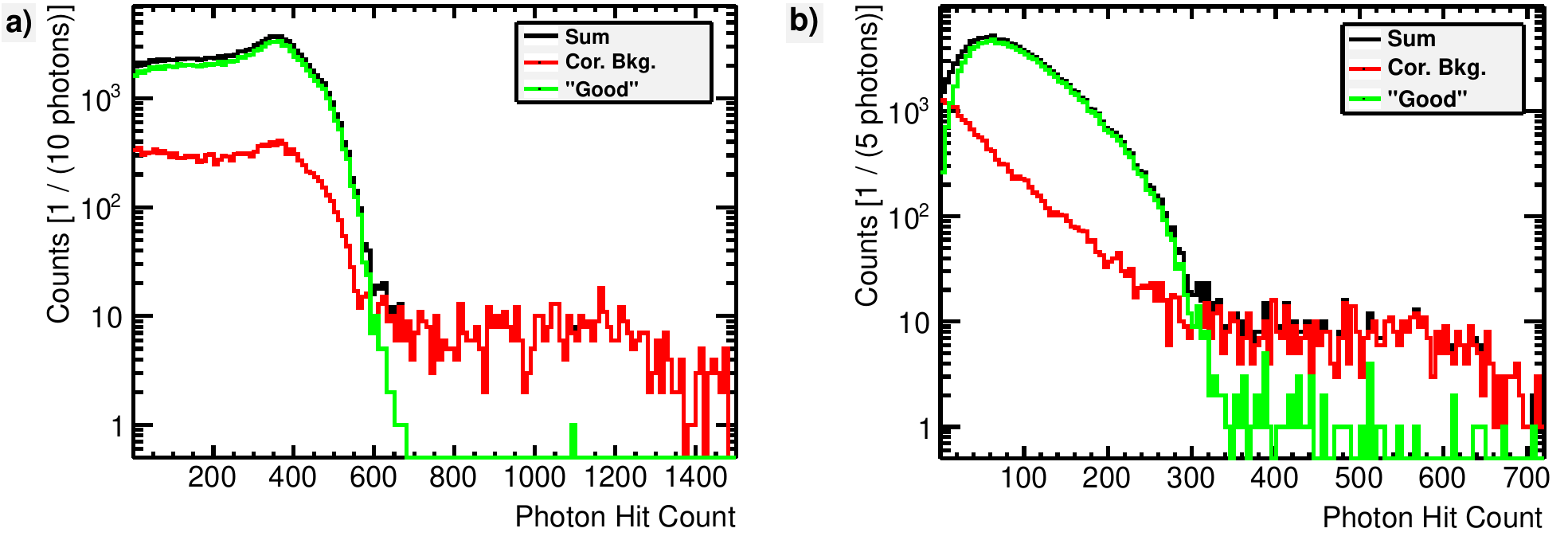} 
 \caption{ Energy spectra for triple coincidences obtained by the Geant4-based Monte Carlo simulation for (a) one outer detector and (b) the central 1'' PMT. Shown are the spectra including all triple coincidences (black), the ``good'' events (green) and correlated background events (red).
}
 \label{fig:spec_sim_dis}
\end{figure*}

The recorded events are analyzed offline. A constant threshold trigger algorithm is utilized to exclude all non-triple coincidences from the subsequent analysis. The energy of an interaction is obtained by integrating the pulse\footnote{The pulse is defined as the range where the waveform is above 10\% of the pulse's maximum amplitude.} appearing in the recorded waveform. Energy spectra for measured events are presented in figure \ref{fig:spec_meas}. Plot (a) shows the energy spectra of the left (continuous blue line) and right (dashed red line) outer detector. The energy spectrum obtained from the 1'' PMT is shown in plot (b). \\
These spectra can be compared to the energy spectra of triple coincidences, obtained from the simulation. These are shown in figure \ref{fig:spec_sim_dis} (a) for one outer detector and in figure \ref{fig:spec_sim_dis} (b) for the 1'' PMT. The distribution including all triple coincidences is drawn in black. \\
``Good'' events correspond to the green spectrum while correlated background events are depicted in red. Different types of correlated background events are identified by the MC-truth-based event classification:
\begin{itemize}
 \item Events, where the positron deposits energy in the LS, but annihilates outside the scintillation detector (outside annihilations). These events are the dominant background for photon hit counts of the 1'' PMT in the range below 100 photons.
 \item Events including a hit of the $1275\,\mathrm{keV}$ gamma in at least one scintillating volume. These events dominate the spectra in the high-energy region (above 600 photons for a 3'' PMT or 340 photons for the 1'' PMT).
 \item Backscatter events, where one single $511\,\mathrm{keV}$ gamma hits both outer detectors. These events are a minor source of correlated background.
 \item Triple coincidences produced by both $511\,\mathrm{keV}$ gammas only. They typically lead to a low energy deposition in the LS volume, but are a less dominant part of the background as compared to outside annihilations.
\end{itemize} 
By comparing figures \ref{fig:spec_meas} and \ref{fig:spec_sim_dis}, the origin of the high-energy interactions (17--30$\,\mathrm{Vns}$ for figure \ref{fig:spec_meas} (a) and 0.44--0.64$\,\mathrm{Vns}$ for \ref{fig:spec_meas} (b) is found to be $1275\,\mathrm{keV}$ gammas. The energy spectra of the outer detectors are dominated by $511\,\mathrm{keV}$ gamma interactions, while most energy depositions in the LS are caused by positrons. ``Good'' events dominate the region of the spectra below the $1275\,\mathrm{keV}$ gamma plateau. \\
Significant qualitative differences exist between measured and simulated spectra. The steep areas in the simulated spectra (around 550 photon hits in figure \ref{fig:spec_sim_dis} (a) and around 250 photon hits in figure \ref{fig:spec_sim_dis} (b)) are not present in the measured ones. This can be explained by effects decreasing the energy resolution of the setup, which are not included in the simulation. For example, statistical uncertainties of the energy determination are caused by the PMTs and the readout electronics are not covered by the simulation. \\
The number of events having a very low energy deposition in the LS (less than $0.1\,\mathrm{Vns}$ in figure \ref{fig:spec_meas} (b) exceeds the simulated prediction. This may be caused by dark noise of the 1'' PMT, which is not implemented in the simulation. \\


The arrival time of pulses in the recorded waveforms is reconstructed via a constant-fraction algorithm. The position, where the signal exceeds 20\% of its maximum amplitude, is used as the reconstructed arrival time. \\
The lifetime of each event is the time difference between the start and the stop signal. Since the stop signal consists of interactions in each of the outer detectors, the mean of both arrival times is used. Hence, the event-based lifetime $\tau$ evaluates as
\begin{equation}
	 \label{eq:lifetime}
\tau = (T_{CF,1} + T_{CF,2})/2 - T_{CF,3}
\end{equation}
where $T_{CF,i}$ denote the signal arrival times, reconstructed via a constant fraction algorithm. The index $i$ takes the values 1, 2 and 3, which refer to the left outer detector, the right outer detector and the central 1'' PMT, respectively. 

The offline analysis was applied to events, which had been recorded during two runs. Run 1 was performed about 1 day after filling the LS vessel. The second data taking (run 2) was done about 35 days after filling in order to investigate the temporal stability of the results. Run 1 (run 2) consists of 16 million (15 million) recorded events, containing 6,817,547 (6,148,912) triple coincidences. \\
Several quality cuts were applied to the data. After all non-triple coincidences are excluded, events exhibiting a small or large energy detected in at least one of the outer detectors are removed. As lower and upper thresholds on the integral of the pulses $2\,\mathrm{Vns}$ and $16\,\mathrm{Vns}$ were chosen. Additionally, events are removed, if the time difference of the two stop signals $\left|T_{CF,1} - T_{CF,2}\right| $ exceeds $0.7\,\mathrm{ns}$. For run 1 (run 2) about 76\% (79\%) of all triple coincidences pass the energy cut and of these about 69\% (68\%) also pass the time difference cut. \\
After these cuts are applied, the data of each run is divided into several groups, depending on the events' energy deposited in the LS (LS-energy). This was done, as the signal-to-background ratio shows a dependence on the energy deposited in the LS, as visible in figure \ref{fig:spec_sim_dis} (b). \\
A fit is performed for each lifetime spectrum, corresponding to one group of events. The fit function used is:
\begin{equation}
	\label{eq:PoLiDe:fit}
f(t) = \left(n_1 \mathrm{e}^{- \frac{t}{\tau_1}} + n_2 \mathrm{e}^{- \frac{t}{\tau_2}}\right) \ast \left(\sum_{i=1}^2 \frac{\eta_i}{\sqrt{2 \pi \sigma_i^2}} \mathrm{e}^{- \frac{(t - \mu)^2}{2 \sigma_i^2}}\right) + n_3
\end{equation}
The two exponential functions in the first bracket correspond to direct annihilations and the decay of p-Ps (index 1), and the decay of o-Ps (index 2). The term within the second bracket is the model used for the time resolution. It is composed of two Gaussian distributions and it is analogous to the model defined in formula \ref{eq:detModel}, except for the fact, that both mean values of the Gaussian distributions have to be equal here\footnote{This was done, as the difference in the means $\mu_i$ of the Gaussian distributions in formula \ref{eq:detModel} was found to be less than $100\,\mathrm{ps}$}. $n_3$ is a constant, describing uncorrelated background. \\
The MINUIT-based analysis package RooFit \cite{roofit}, included in the ROOT framework \cite{root} is used to perform $\chi^2$ fits. \\

\begin{figure}[t]
 \centering
 \includegraphics[width=8.8cm]{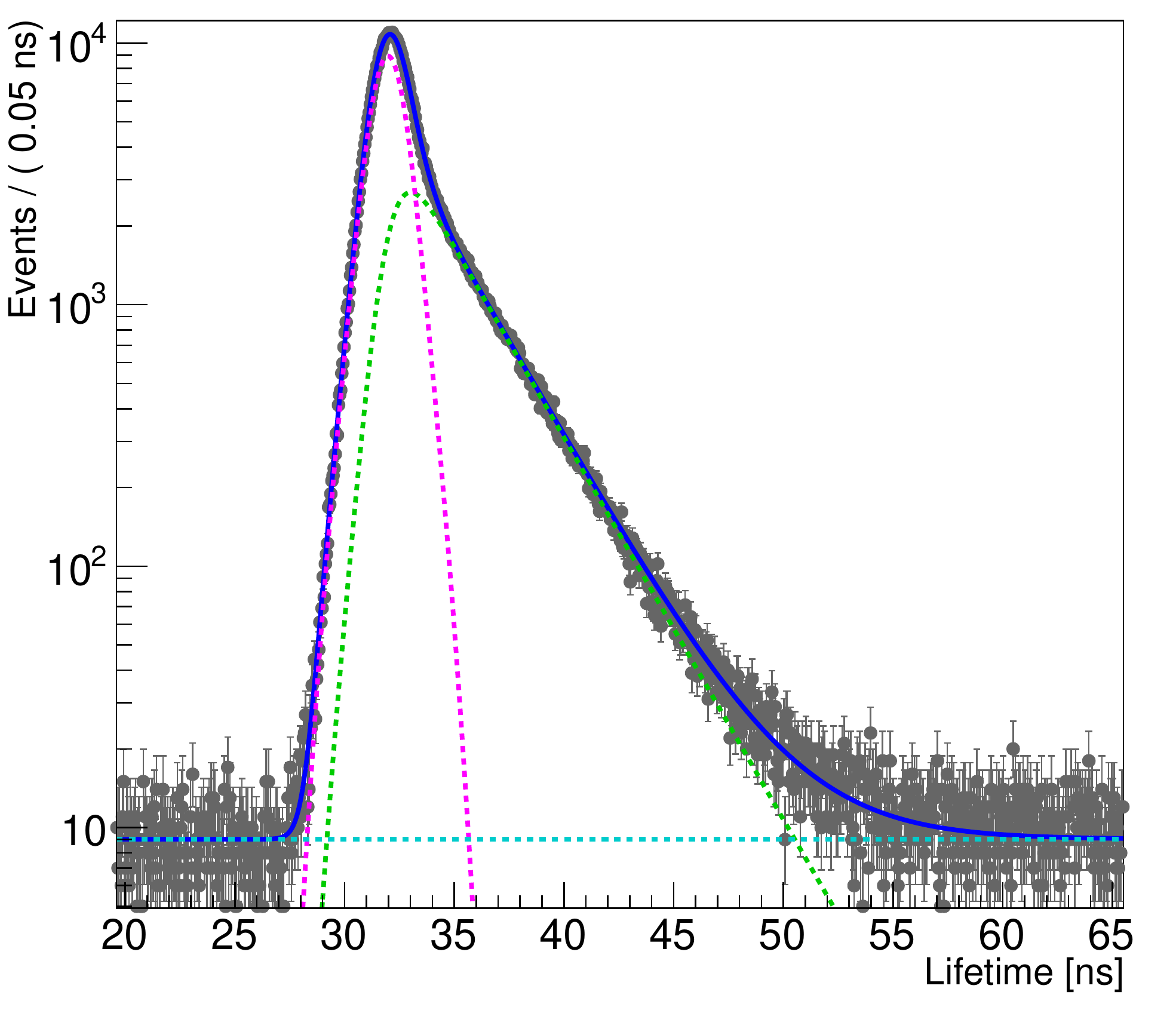}
 \caption{ A lifetime spectrum of events from run 1, consisting of 634,861 events. These events pass all quality cuts (see text) and the energy deposition in the liquid scintillator is within $0.1\,\mathrm{Vns}$ and $0.15\,\mathrm{Vns}$. The energy is reconstructed by integrating the pulse, appearing in the waveform of the central 1'' PMT. The continous line is the result of a fit, using formula \ref{eq:PoLiDe:fit}. Dashed lines indicate components of the fit function (color online): short-lived component (magenta), long-lived component (decaying orthopositronium, green) and a constant term (cyan).}
 \label{fig:lifetime}
\end{figure}

Figure \ref{fig:lifetime} shows a lifetime plot. The events from run 1 were used for this histogram and the energy release of these events in the LS is within $0.1\,\mathrm{Vns}$ and $0.15\,\mathrm{Vns}$ (pulse integral). A fit of this distribution was performed, using formula \ref{eq:PoLiDe:fit}. The continuous line is the result of this fit and the dashed lines represent components of the fit function: The magenta and green dashed lines refer to the exponentials with index 1 and 2, respectively. The horizontal cyan dashed line indicates uncorrelated background. \\
For both runs fits were performed in the LS-energy range from (0.05 -- 0.45)$\,\mathrm{Vns}$. The range was divided into eight sections, each  having a width of $0.05\,\mathrm{Vns}$. The resulting 16 event groups were investigated separately. The reduced $\chi^2$ values were in the range between 0.89 and 1.11, except for the energy region of (0.05 -- 0.10)$\,\mathrm{Vns}$, where the reduced $\chi^2$ value is 4.67 (5.85) for run 1 (run 2) and the energy region of (0.4 -- 0.45)$\,\mathrm{Vns}$, for which a normalized $\chi^2$ value of 0.78 (0.73) is obtained for run 1 (run 2). \\
The detector resolution is described by $\sigma_1$, $\sigma_2$ and $\eta_1$ from formula \ref{eq:PoLiDe:fit}, which were fitted as free parameters. Values of $\sigma_1$ are in the range of (0.46 -- 1.0)$\,\mathrm{ns}$. Lower values of $\sigma_1$ correspond to higher energy depositions in the LS, in agreement with the expectation that a higher energy release increases the number of produced photons, which enhances the resolution of the reconstruction of the start signal. Values of $\sigma_2$ are in the range of (0.8 -- 1.7)$\,\mathrm{ns}$ for energies between $0.10\,\mathrm{Vns}$ and $0.45\,\mathrm{Vns}$, without any observable dependence\footnote{In the energy region between $0.05\,\mathrm{Vns}$ and $0.10\,\mathrm{Vns}$, $\sigma_2$ is about $4.2\,\mathrm{ns}$ for both runs.}. The fraction of the narrow component of the resolution $\eta_1$ varies in a range of 0.46 and 0.96. The deviations from the simulated vaules can be explained by background, which is not covered in the simulation of the time resolution of the detector. \\

\begin{table}[t]
\begin{center}
	\caption{Results for the lifetime $\tau_2$ and formation probability $I_2$ of orthopositronium in the investigated liquid scintillator (LS). Each of the rows corresponds to events selected according to the energy deposition in the LS (LS-energy). Run 1 (1 day after filling the LS vessel) was used here.}
	\begin{tabular}{|c|cc|} 
	\hline  LS-energy range [Vns] & $\tau_2$ [ns] & $I_2$ [\%]  \\ 
	\hline
	0.05 -- 0.10 & $2.887\pm0.009$ & $36.8\pm0.3$	\\
	0.10 -- 0.15 & $2.973\pm0.009$ & $42.0\pm0.3$  \\
	0.15 -- 0.20 & $2.951\pm0.012$ & $44.4\pm0.4$  \\
	0.20 -- 0.25 & $2.98\pm0.02$ & $47.4\pm0.4$  \\
	0.25 -- 0.30 & $3.12\pm0.04$ & $47.9\pm0.6$  \\
	0.30 -- 0.35 & $3.27\pm0.07$ & $45.5\pm0.9$  \\
	0.35 -- 0.40 & $3.66\pm0.15$ & $41.6\pm1.5$  \\
	0.40 -- 0.45 & $3.8\pm0.3$ & $34\pm2$  \\
	\hline
	\end{tabular} 
	\label{tab:results:run1}
\end{center}
\end{table}

\begin{table}[t]
\begin{center}
	\caption{As table \ref{tab:results:run1}, for run 2 (35 days after filling the LS vessel).}
	\begin{tabular}{|c|cc|} 
	\hline  LS-energy range [Vns] & $\tau_2$ [ns] & $I_2$ [\%]  \\ 
	\hline
	0.05 -- 0.10 & $2.917\pm0.010$ & $34.3\pm0.3$  \\
	0.10 -- 0.15 & $2.998\pm0.010$ & $40.8\pm0.3$  \\
	0.15 -- 0.20 & $2.943\pm0.012$ & $43.5\pm0.3$  \\
	0.20 -- 0.25 & $2.941\pm0.019$ & $47.5\pm0.5$  \\
	0.25 -- 0.30 & $3.04\pm0.03$ & $48.4\pm0.5$  \\
	0.30 -- 0.35 & $3.20\pm0.06$ & $46.4\pm0.7$  \\
	0.35 -- 0.40 & $3.29\pm0.09$ & $44.7\pm1.0$  \\
	0.40 -- 0.45 & $3.62\pm0.16$ & $37\pm2$  \\
	\hline
	\end{tabular} 
	\label{tab:results:run2}
\end{center}
\end{table}

\begin{figure}[t]
 \centering
 \includegraphics[width=7cm]{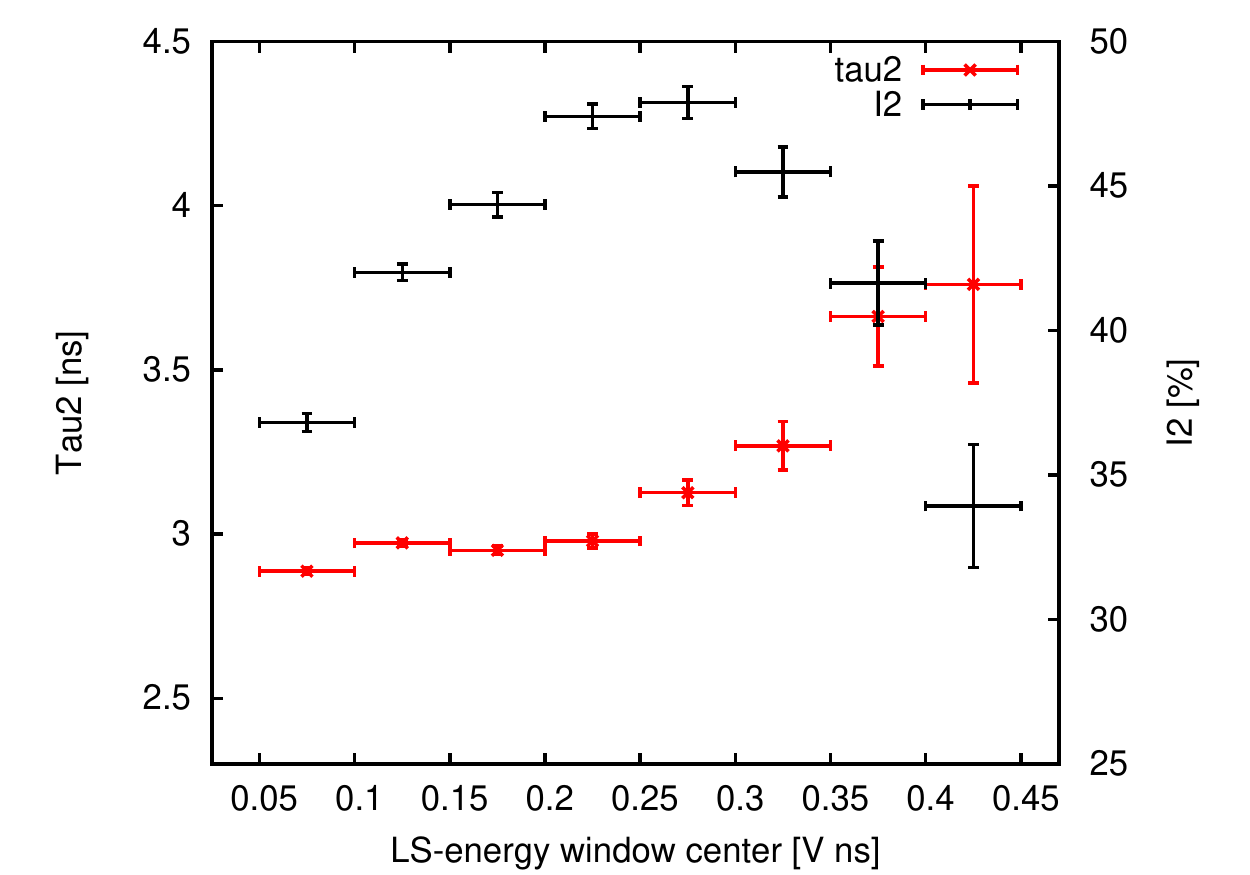}
 \caption{ Lifetime $\tau_2$ and formation fraction $I_2$ of orthopositronium as a function of the energy deposition in the liquid scintillator (LS-energy). The values correspond to fit results using run 1 data (see table \ref{tab:results:run1}).}
 \label{fig:results:run1}
\end{figure}

\begin{figure}[t]
 \centering
 \includegraphics[width=7cm]{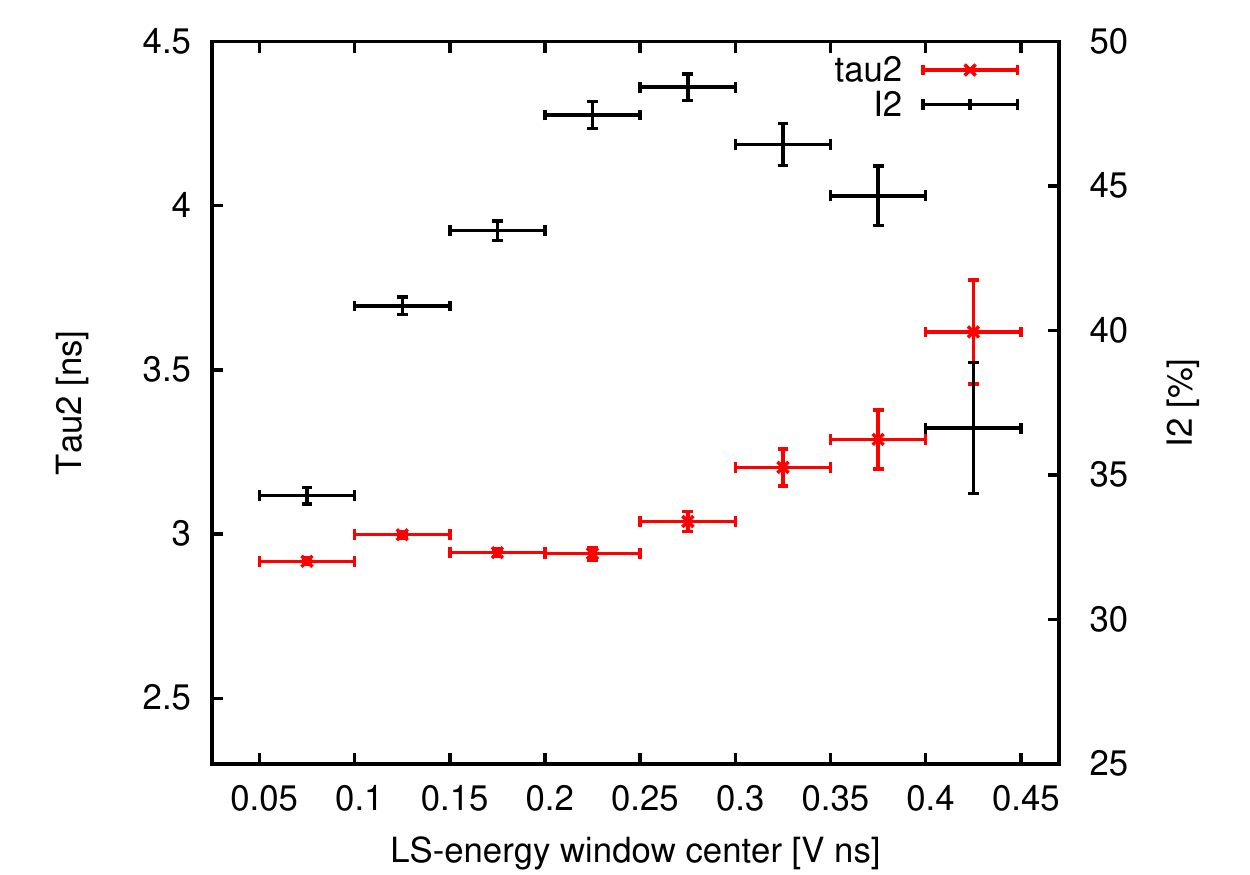}
 \caption{ As figure \ref{fig:results:run1}, using run 2 data (see table \ref{tab:results:run2}).}
 \label{fig:results:run2}
\end{figure}

Values of $\tau_2$ and $I_2$, corresponding to different energy regions are presented in table \ref{tab:results:run1} and \ref{tab:results:run2} for run 1 and run 2, respectively. Plots are shown in figure \ref{fig:results:run1} and \ref{fig:results:run2}. The presented uncertainties represent the statistical uncertainty, obtained from the fits. The lifetimes of o-Ps are not strong\-ly dependent on the LS-energy region; the deviations at high LS-energies can be attributed to effects caused by the increasing influence of $1275\,\mathrm{keV}$ gamma background. No significant LS-energy depencence of $\tau_2$ is expected from theory \cite{2002JChPh.116.6178S}. No significant differnces between run 1 and 2 were found. \\
The final $\tau_2$ value and its uncertainty are obtained from the arithmetic average of the individual results, weighted by using their statistical uncertainties. Since the background contribution is low in the LS-energy region of (0.1 -- 0.35)$\,\mathrm{Vns}$ according to the simulation, only results in this regions are used for averaging. We obtain $\tau_2 = 2.97\,\mathrm{ns} \pm 0.04\,\mathrm{ns}$ by combining the results of both runs. \\
The obtained values of $I_2$ are more dependent on the LS-energy than $\tau_2$. $I_2$ is maximal for LS-energies around (0.25 -- 0.30)$\,\mathrm{Vns}$ and decreases both with increasing and decreasing energies. This behaviour is similar for both runs. Several possible explanations are discussed in the following:
\begin{itemize}
 \item It can be assumed that most correlated background events passing the quality cuts, appear in the short lifetime component of the lifetime spectrum. Firstly, correlated background events only involving gamma particles are not expected to result in a delay of the order of $\mathrm{ns}$ between start and stop signal. Secondly, if a positron escapes from the LS after depositing some of its kinetic energy there, it will most likely hit the aluminum wall of the vessel, resulting in a lifetime of the order of $100\,\mathrm{ps}$ \cite{bell:graham}. Hence, one can expect that correlated background events lead to a decrease of the apparent o-Ps formation probability. \\
 This effect was studied using the simulated background composition. Assuming, that all background events appear in the short lifetime component of the lifetime spectrum and the ``true'' o-Ps formation fraction\footnote{This number was chosen such that simulated and measured $I_2$ vaulues are equal for the LS-energy, where the measures $I_2$ value is maximal.} was $I_{2,true} = 50.7\%$, the simulated expectation of $I_2$ was obtained, which is presented in figure \ref{fig:compare} (red diagonal crosses). The figure also shows $I_2$, produced by fits of run 1 (see table \ref{tab:results:run1}) for comparison. \\
 The results of the simulation are in agreement with the fit results for LS-energies of at least $0.2\,\mathrm{Vns}$, while the $I_2$ values produced by the fits are falling below the simulated values significantly for lower LS-energies. Additional background events, which are not accounted for in the simulation, could explain this discrepancy. \\
 An alternative interpretation is based on the fact, that the LS-energy reconstruction of the measurement has a lower resolution than the energy reconstruction of simulated events. Hence, events from the low-LS-energy region exhibiting a high proportion of background are shifted towards higher energies. \\
 The following aspects could also explain the discussed discrepancy:

\begin{figure}[t]
 \centering
 \includegraphics[width=7cm]{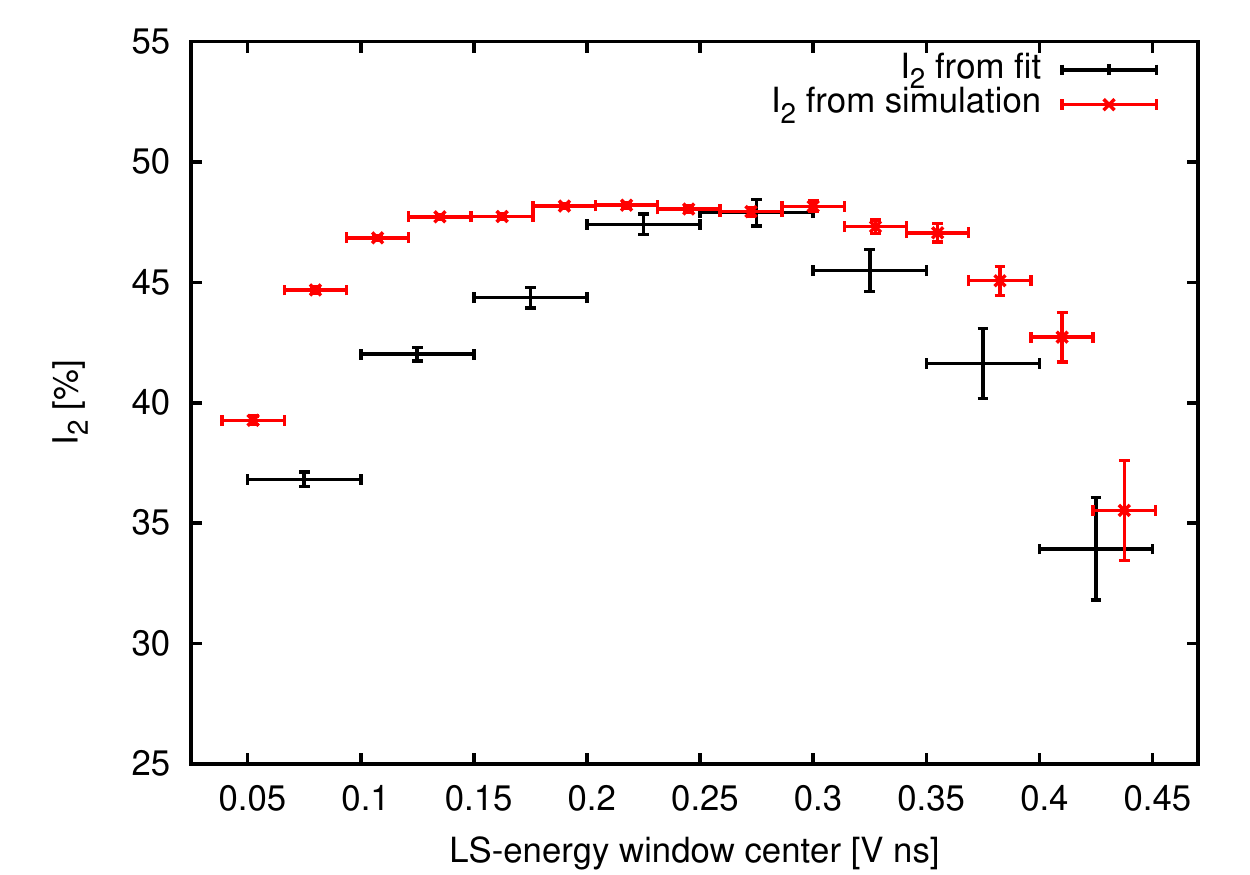}
 \caption{ Formation fraction $I_2$ of orthopositronium as a function of the energy deposition in the liquid scintillator (LS-energy). The black data points correspond to the result of fits using run 1 data (see table \ref{tab:results:run1}). The red data points (diagonal crosses) are the result of the Geant4-based simulation, estimating the influence of correlated background on $I_2$. }
 \label{fig:compare}
\end{figure}

 \item The decrease of $I_2$ for low LS-energies could be due to the decreasing time resolution, which may affect the reconstruction of $I_2$ in fits of lifetime distributions. 

 \item Another aspect, which may cause the decrease of $I_2$ for low LS-energies is, that positrons with higher kinetic energies are expected to stop and form Ps deeper in the LS than positrons with lower energies. Since the sample is a mixture of different molecules, demixing is possible. Thus, the composition of the surrounding medium of stopped positrons is dependent on their initial kinetic energy. The dependence of $I_2$ on the chain lengths of different alkylbenzenes is known \cite{gcs}. Longer alkyl chains tend to increase $I_2$, as shown for chains consisting of 1, 2, 4, 5 and 8 carbon atoms \cite{gcs}. $I_2$ ranges from $(30.1\pm2.3)\%$ (ethylbenzene) to $(40.1\pm5.1)\%$ (n-octylbenzene) for these molecules. However, since the densities of alkylbenzenes are decreasing and the boiling points are increasing for longer chain lengthes \cite{CRChandbook:2014}, one would expect an increased concentration of alkylbenzenes with a longer chain length in the upper part. Thus, one would expect $I_2$ to increase for decreasing penetration depths of the positron (decreasing LS-ener\-gies), which is in conflict with our measurement results. \\

 \item Finally, the formation probability of Ps (and hence o-Ps) in a liquid depends on the kinetic energy of the positron, according to the \textit{blob} model, presented in \cite{2002JChPh.116.6178S}. This is because positrons in liquids have to catch thermalized free electrons when forming Ps. These electrons are the product of ionization caused by the positron during its deceleration. However, since the positron catches an electron predominantly from the terminal part of its track (the so-called \textit{blob}), where the positron's kinetic energy is of the order of $500\,\mathrm{eV}$, the differences in $I_2$ are expected to be negligible for positron energies in the range of $100\,\mathrm{keV}$. 

\end{itemize}

Since the process causing the decrease of $I_2$ in the energy region between $0.05\,\mathrm{Vns}$ and $0.20\,\mathrm{Vns}$ could not be identified, no final result for the formation probability of o-Ps in the LAB-based LS of JUNO is presented here. However, if we assume that correlated background events are causing the deviations, the highest obtained $I_2$ value, about $48\%$, can be interpreted as a lower limit for the true formation probability of o-Ps. \\

\section{Discussion}

The measured lifetime of o-Ps $\tau_2 = 2.97\,\mathrm{ns} \pm 0.04\,\mathrm{ns}$ can be compared to literature values for $\tau_2$ in pure LAB. Our result is slightly lower than both results $\tau_2 = 3.08\,\mathrm{ns} \pm 0.03\,\mathrm{ns}$ measured by \cite{franco} and $\tau_2 = 3.10\,\mathrm{ns} \pm 0.07\,\mathrm{ns}$ measured by \cite{Cheng:2016ego}. This deviation can be attributed to the effect of wavelength shifters present in our LS sample. \\
A discrepancy exists between $I_2$ values for pure LAB, found in the literature. $I_2 = 54.2\% \pm 0.5\%$ is reported by \cite{franco}, which is significantly higher, than $I_2 = 43.7\% \pm 1.2\%$, found by \cite{Cheng:2016ego}. This discrepancy may be caused by the PALS setup, used by the mentioned experiments. As explained earlier, in PALS measurements positron signals are recorded regardless of the medium where the positron interacts. Both mentioned experiments use different techniques to account for positrons interacing within the source or its encapsulation. \\ 
We are not able to give a final result for $I_2$. The lower limit of $I_2 \gtrsim 48\%$ presented in the previous section is conflicting with the value found by \cite{Cheng:2016ego}.




\section{Conclusion}

The formation of o-Ps in organic LS based neutrino experiments induces a delay between the signals caused by the deposition of the positron's kinetic energy and the annihilation gamma interaction. Since the lifetimes of o-Ps in LS are of the order of $3\,\mathrm{ns}$, the detector pulse shapes are distorted, which affects reconstruction algorithms. On the other hand a discrimination between electron and positron induced events is possible. The most important quantities for these aspects are the lifetime $\tau_2$ and the formation probability $I_2$ of o-Ps. \\
In this work we measured $\tau_2$ for the full LAB-based liquid scintillator of JUNO, including both wavelength shifters. In contrast to the PALS setup used by previous experiments, we used a novel setup type, which allows to suppress events, where the positron interacts outside of the LS sample. \\
The setup was characterized and optimized using a Geant4-based Monte Carlo simulation. Additionally, the time resolution of our setup was simulated, based on single-photon transit time measurements, performed for each of the three PMTs used. Values for $\tau_2$ and $I_2$ are produced by a fit of lifetime spectra. Since a significant dependence of $I_2$ on the energy released in the LS (LS-energy) was observed, the fits were repeated for different LS-energy regions. A final value $\tau_2 = 2.97\,\mathrm{ns} \pm 0.04\,\mathrm{ns}$ is obtained. However, since $I_2$ is dependent on the LS-energy for reasons not completely understood, no final $I_2$ value is given here. Under the assumption that the dependency is caused by background, we can give a lower limit $I_2 \gtrsim 48\%$. \\


\section*{Acknowledgements}

This work was supported by the DFG Cluster of Excellence ”Origin and Structure of the Universe”, the DFG research unit ”JUNO” and the Maier-Leibniz-Laboratorium in Garching.


\bibliography{mybibfile}

\end{document}